\documentclass[runningheads,a4paper]{llncs}

\usepackage{amssymb}
\usepackage{graphicx}

\usepackage{url}
\urldef{\mailsa}\path|{alfred.hofmann, ursula.barth, ingrid.haas, frank.holzwarth,|
\urldef{\mailsb}\path|anna.kramer, leonie.kunz, christine.reiss, nicole.sator,|
\urldef{\mailsc}\path|erika.siebert-cole, peter.strasser, lncs}@springer.com|

\begin{document}

\mainmatter  

\title{Evolution of Cosmological Total Energy Density and Transient Periods in Cosmology}

\titlerunning{Total Energy Density and Transient Periods in Cosmology}

%
%
\author{Bob Osano%
}
\authorrunning{Total Energy Density and Transient Periods in Cosmology}

\institute{Cosmology and Gravity Group, Department of Mathematics and Applied Mathematics\\$\&$\\
Centre for Higher Education Development, \\University of Cape Town (UCT), Rondebosch 7701, Cape Town, South Africa\\
}

\toctitle{Cosmology}
\maketitle

\begin{abstract}
The evolution of the Universe is traditionally examined by monitoring how its material content evolves as it expands. This model of an isolated system is expressed as the equation of motion of the bulk but segmented into different epochs. In particular, the evolution of the {\it $Friedmann-Lema\hat{i}tre-Robertson - Walker$} (FRLW) Universe is separated into different epochs that are characterised by the dynamics of whichever mass-energy density constituent is dominant at the time. The standard analysis of the evolution of the Universe in a particular epoch often considers the evolution of the dominant energy density only, disregarding all others. Whereas this represents the limiting case, in principle the contributions from others cannot always be ignored particularly in the vicinity of the equality of the various competing mass-energy densities or the transition periods between epochs. We examine the evolution of the total energy density rather than individual energy densities during the different epochs. We find that taking into account the contributions from the various constituents leads to a broader range of possible evolution histories which enriches the standard picture. 
\end{abstract}

\section{Introduction}
How well do we understand the evolution of the universe? The answer to this question depends on which cosmologist you speak to. Needless to say, the recent discovery of an accelerated expansion, \cite{Perlmutter98}, is contrary to expectation and is forcing cosmologists to reexamine methodologies and approaches used in this field of study. This cosmic acceleration demands the reexamination of both the mathematical and physical assumptions employed in the modelling processes \cite{Frieman08} and the reevaluation of observational data analysis. At the core of the issue, is what material substance makes up the universe and how the evolution of such material substance affects the evolution of the universe as a whole \cite{Mortonson13}. Until cosmic acceleration was discovered, the role played by the relativistic and the non-relativistic matter in the universe's expansion rate was thought to be well understood and accounted for in the standard $\Lambda CDM$ model. 

The densities of the material making up the universe have different rates of decay as the universe expands. These evolve in either a decoupled or a coupled manner. The significance of the different rates of evolution is that the universe undergoes phases where one substance plays the dominant role, thereby driving the evolutionary history of the whole universe for a period. The different decay rates mean that there are points in time when there is a switch in the dominant material. The dynamics of the various known materials have been investigated and are well documented. These inform our current understanding of the Universe's evolutionary history \cite{Ray,Zelik}. For instance, the universe underwent periods when the dominant material switched from radiation to matter. In particular, it is accepted that when the early universe was $\approx$ 47,000 years in cosmic time or about $z=3600$, the energy density of matter became greater than the radiation energy density. However, photons could not freely stream as the Universe was optically thick for a considerable period. This is thought to have lasted until the Universe was about 378,000 years old ($z= 1100$). For this reason, the radiation-dominated era which in theory ends at approximately  $z=1100$, in effect last until about $z=3600$. This transient period suggests that the dynamics of both matter and radiation are important in our understanding of the evolution of the Universe in the protracted period. We could ask the question, what information is lost when an instantaneous switch of domination is preferred, as is done in literature, to the gradual change?  Pointedly, what can we learn from considering a gradual transition between epochs?  

The chapter is organised as follows: Section (\ref{one}) reviews Friedmann equations. Section (\ref{two}) discusses the mass-energy composition of the universe and how the total evolves. The effect of the interacting dark sector on the evolution of the total energy density is modelled in section (\ref{three}) and the various epochs are considered. Discussions and conclusions are found in section (\ref{DC}).

\section{\label{one}Friedmann Equations}
The hot big-bang cosmological model (the reader is referred to \cite{Wein0,Peebles93}) is presently the preferred model of the universe. This is a model built on the isotropic and homogeneous $Friedmann-Lema\hat{i}tre-Robertson-Walker$ (FLRW) solution of Einstein's equation from general relativity, where the expansion of the Universe is manifested in the cosmic scale factor $a(t)$\cite{Shu}. The evolution of the scale factor is governed by the Friedmann equations which take the form:
\begin{eqnarray}\label{Frid1}
\frac{\dot{a}^2}{a^2}&=&\frac{8\pi G}{3}\rho_{Tot}-\frac{k}{a^2}+\frac{\Lambda}{3},\\
\frac{\ddot{a}}{a}&=&-\frac{4\pi G}{3} (p_{Tot}+3p_{Tot}-\frac{\Lambda}{3},
\end{eqnarray}
where we have set the speed of light to $c=1$. We note that the total energy density $\rho_{Tot}$ does not include the curvature or the cosmological constant term. Now since the Hubble parameter $H=\dot{a}/a$, equation (\ref{Frid1}) can be appropriately normalised to read
\begin{eqnarray}\label{eqn1}
1 &=&\frac{8\pi G}{3H^2}\rho_{Tot}-\frac{k}{a^2H^2}+\frac{\Lambda}{3H^2},
\end{eqnarray} provides a simple yet effective way of discussing the energy-density composition of the universe.
\section{\label{two}The evolution of total energy density}
Let us represent the total density of the universe at a particular time in its cosmic history by $\rho_{Tot}$. This total has contributions that include radiation (relativistic particles), baryonic matter, non-baryonic matter, and other types of energy densities.
For ease of reference, we will use the following notations to denote these constituents of the total energy density: namely $\rho_{r}$ (radiation), $\rho_{nb}$(non-baryonic which will later be identified as dark-matter ), $\rho_{b}$(ordinary or baryonic matter), $\rho_{o}$ (other forms of energy density which will later be identified as dynamical dark energy)\cite{Peeb} and $\rho_{\Lambda}$ (the cosmological constant). We have not made any assumption about dark energy and cosmological constant as it may be impossible to distinguish cosmological constant and vacuum energy, see \cite{Ruth}. Dark energy or vacuum energy is a form of energy that is postulated to be responsible for the observed late-time accelerated expansion of the Universe \cite{Riess}. The total mass-energy density is therefore given,\begin{eqnarray}\rho_{Tot}=\sum_{i=r, b, nb, o}\rho_{i}
\end{eqnarray} The analysis of how the universe evolves takes into account the fact that at different periods different energy densities dominate. To determine what is dominant, it is customary to compare the ratios of each contributor to the total.

\begin{eqnarray}\label{norm1}1=\frac{1}{\rho_{Tot}}\sum_{i=r, b, nb, o}\rho_{i},
\end{eqnarray} This is not to be confused with the ratio $\rho_{r}/\rho_{c}$ where $\rho_{c}=3H^2/8\pi G$ given $\kappa=0$.\footnotetext{We note that the critical-density ($\omega_{crit} = 1$) implies a flat Universe, the sub-critical density ($\omega < 1$) implies a negatively curved ($\omega> 1)$ is positively curved Universe.} Equation (\ref{eqn1}) may be formulated in terms of the various energy densities as follows:
\begin{eqnarray}\label{eqn2}
1 &=&\frac{8\pi G}{3H^2}\rho_{Tot}\left(\frac{1}{\rho_{Tot}}\sum_{i=r, b, nb, o, \Lambda}\rho_{i} \right)=\left(\frac{1}{\rho_{c}}\sum_{i=r, b, nb, o}\rho_{i} \right)
\nonumber\\
&=&\Omega_{r}+\Omega_{b}+\Omega_{nb}+\Omega_{o}+\Omega_{\Lambda},
\end{eqnarray} where $\Omega$ is the density parameter and is defined as the ratio of the observed to the critical density $\rho_{c}$. In terms of observation, the parameters are normalised to agree with observations today. In particular, a density parameter is multiplied with the square of the Hubble constant $h$; for example, $\Omega_{b}h^2$ and for critical density $\Omega_{c}h^2.$ 

The Hubble constant holds the key to understanding important information about the evolutionary history of the universe and measuring accurately is therefore important. To this end, the exact value of the Hubble constant is still unknown. This is because different results are obtained depending on how where one starts For example from fundamental physics, which is thought to have driven the evolution of the universe, the Hubble constant is calculated to be about 68 km/s/Mpc. While what is found from observations of different cosmic objects ranges from 69.8 to 74 68 km/s/Mpc \cite{Hin,Agh,Rie,Bir}. It is not at present if the disagreement between the calculated and the measured value points to the need for new physics; a subject that is of current debate \cite{Kno} or a rethink of the modeling process. In this chapter, we examine how incorporating the constituent density ratios affects the growth of the total energy density. 
                           The equation gives the case where no new physics or non-standard interaction is needed,
\begin{eqnarray}
\rho_{c}\equiv\rho_{Tot}. 
\end{eqnarray} Contemplate, for the moment, a situation where equation (\ref{eqn2}) does not hold. There are several plausible scenarios for this, for example where a previously unknown and initially in-efficacious material evolves to the point that its contributions become sufficiently significant to start impacting the dynamics of the model. This seemingly contrived idea, speaks to the current theories of dark energy and dark matter. The impact of this is that theories build on the current understanding of energy balance leading to the current $\rho_{c}$ needing reexamination. In particular, the consequence of $\rho_{c}<\rho_{Tot}$ or $\rho_{c}>\rho_{Tot}$ to the evolution history of the model would need to be investigated. What if the universe evolves to the point where conditions allow previously non-interacting material to interact thereby altering the density balance briefly?  
Let us now create a generic template for discussing these what-if scenarios. If we denote these ratios as follows \begin{eqnarray}\label{eqndef}\rho_{r}/\rho_{Tot}&=&\alpha, \rho_{b}/\rho_{Tot}=\beta,\\\rho_{nb}/\rho_{Tot}&=&\gamma,\rho_{o}/\rho_{Tot}=\delta,\rho_{\Lambda}/\rho_{Tot}=\xi\end{eqnarray}then the sum is
\begin {eqnarray}\label{eqratio00}
\alpha+\beta+\gamma+\delta+\xi=1.
\label{sum}\end{eqnarray}These are cosmic-time depended ratios, i.e. $\alpha=\alpha(t)$, where 
\begin {eqnarray}\label{eqratio01}
\dot{\alpha}\ne0\ne\dot{\beta}\ne0\ne\dot{\gamma}\ne0\ne\dot{\delta}\ne0&\ne&\dot{\xi},\end{eqnarray} 
their derivatives taken concerning cosmic time are non-vanishing.
\begin{eqnarray}
\dot{\alpha}+\dot{\beta}+\dot{\gamma}+\dot{\delta}+\dot{\xi}=0.
\end{eqnarray}

This means that ratios evolve in such a way that the sum of the time derivatives of the ratios will always have zero. The crucial takeaway from this discussion is that densities evolve so that the sum of these fractions remains a unit. In the present analysis, we will assume that non-baryonic matter makes up the dark matter ($DM$) and baryonic matter is the normal matter, represented by the subscript ($m$). We will replace the {\it other} energy density with dynamic dark energy (DDE). If we replace the cosmological constant, $\Lambda$, with non-dynamic dark energy (NDE) then, then it follows that dark energy is a sum of the two.
\section{\label{three}Interactions and evolution of total energy density}
In this section we examine the evolution of the total energy density of a FLRW universe made up of the following: radiation ($\rho_{r}$), baryonic matter ($\rho_{m}$), dark matter ($\rho_{DM}$), dynamical dark-energy ($\rho_{DDE}$) and non-dynamical dark-energy ($\rho_{NDE}$), this distinction is important as what is collectively referred to as dark energy may be made up of distinctive parts. The question to ask is what kinds of interactions take place and are these significant enough to affect how these densities evolve? Consider radiation and matter interaction. Radiation and matter interaction involves ionisation. Radioactive particles or electromagnetic waves with sufficient energy collide with electrons on the atom to knock electrons off the atom. For partially ionised matter, the growth pattern of radiation, ionised matter, and neutral matter will differ from that of just radiation and matter. But is this sufficient to affect the evolution of $\rho_{Tot}$
In this study, we let the dark-sector constituents mimic perfect fluid, which can then be quantified by the usual energy-momentum tensor of the form: $T^{\nu\mu}=(\rho+p)u_{\nu}u_{\mu}+pg_{\nu\mu}$ that obeys the conservation law $\nabla^{\nu}T_{\nu\mu}=0,$where $\rho$ is the energy density, $p$ is the pressure, $u_{\nu}$ is the 4-velocity. We will comment on the general non-perfect fluid case later. The conservation law, for the present case, yields:
 \begin{eqnarray}\label{rhodots1}
\dot{\rho}_{r}&=&-\Theta (1+\omega_{r})\rho_{r}\\\label{rhodots2}
\dot{\rho}_{m}&=&-\Theta (1+\omega_{m})\rho_{m}\\\label{rhodots3}
\dot{\rho}_{DM}&=&-\Theta(1+\omega_{DM})\rho_{DM}+Q_{ME}\\\label{rhodots4}
\dot{\rho}_{DDE}&=&-\Theta(1+\omega_{DDE})\rho_{DDE}-Q_{ME}.\\\label{rhodots5}
\dot{\rho}_{NDE}&=&0,
\end{eqnarray} where $\Theta$ is the expansion and $Q$ is the coupling term between dark matter and dynamical dark energy. Chaplygin\cite{Adams} gas is an example of a model that mimics the behaviours of ${\rho}_{DM}$ and ${\rho}_{DDE}$ in that early and late universe, respectively. 

The barotropic equation of state assumption is not crucial and can be relaxed if a broader picture is required. The coupling of dark matter to dark energy could have a profound effect on structure formation and the evolution thereof, for example, it has been demonstrated in \cite{Fin} that a momentum coupling of the dark sector could lead to a suppression of structure formation, which can be shown by examining the evolution of density fluctuations. We also note that $\omega_{DDE}=-1$ implies $\dot{\rho}_{DDE}=0=\dot{\rho}_{NDE}$ which renders $DDE$ indistinguishable from $NDE$ as pointed out in \cite{Ruth}. But let us consider the evolution of the total mass-energy density. It follows that
\begin{eqnarray}\label{rhoDotTot0}
\dot{\rho}_{Tot}&=&\sum_{i=r, m, DM, DDE, NDE}\dot{\rho}_{i}
\end{eqnarray} Using equations (\ref{rhodots1}-\ref{rhodots5}), we find
\begin{eqnarray}\label{rhoDotTot}
\dot{\rho}_{Tot}&=&-\theta\left(\frac{4}{3}\alpha+\beta+\gamma(1+\omega_{DM})+\delta(1+\omega_{DDE})\right)\rho_{Tot},\nonumber\\
\end{eqnarray} where we have assumed that $m$ is 'dust' whose equation of state is $\omega_{m}=0,$ while radiation's equation of state is $\omega_{r}=1/3$. If we suspend the condition given in equation (\ref{eqndef}) i.e. if we assume that the fractional quantities are momentarily invariant to time then, it follows that the generic form of the total energy density scales as
\begin{eqnarray}\label{eqnSol}
\rho_{Tot} &\propto& { a^{-{3(\frac{4}{3}\alpha+\beta+\gamma(1+\omega_{DM})+\delta(1+\omega_{DDE}))}}}.
\end{eqnarray} This expresses the total density in terms of EoS and the fraction of each constituent. It is clear that for pure radiation $\alpha=1$ and hence $\rho_{Tot}=\rho_{r}=C_{r}a^{-4} $. The case of pure matter is given by setting $\beta= 1$ so that $\rho_{Tot}=\rho_{m}=C_{m}a^{-3} $. Whereas the cosmological constant remains constant, the different evolution patterns of the other constituents ensure that the ratios change over time. By fixing the ratios, we can obtain the evolution equation for the total energy density in each epoch. The simplest case is obtained by using the following ansatz for the equation of state of the dark-sector constituents; $\omega_{DM}=0$ and $\omega_{DDE}=-1$. Interacting dark sector's contribution to the evolution of the total energy density, although hidden in these equations, has the potential to modify the standard picture. The complexity is not aided when $\omega_{DDE}\approx-1$. In the next sections, we will consider different epochs and how the total energy density evolves in such epochs. The more interesting case is where conditions given in equations (\ref{eqratio00}) and (\ref{eqratio01}) hold. As shown in section \ref{App1}, one can demonstrate that the generic solution in this case takes the form;
\begin{eqnarray}\label{rho1}
\rho_{Tot}&=&a^{-3f}e^{3\int (ln a)\dot{
\zeta}dt}\end{eqnarray} where \begin{eqnarray}
\zeta&=&\frac{4}{3}\alpha+\beta+\gamma(1+\omega_{DM})+\delta(1+\omega_{DDE}).
\end{eqnarray}
Equation (\ref{eqnSol}) is recovered when $\dot{\zeta}=0.$ This is a well-studied case and documented case. The more interesting case that we consider in this chapter is that for $\dot{\zeta}\ne0.$ 
 
 \section{\label{Epochq}Epochs and transitions periods}
\begin{figure}[ht!]
\centering
\includegraphics[scale=0.55]{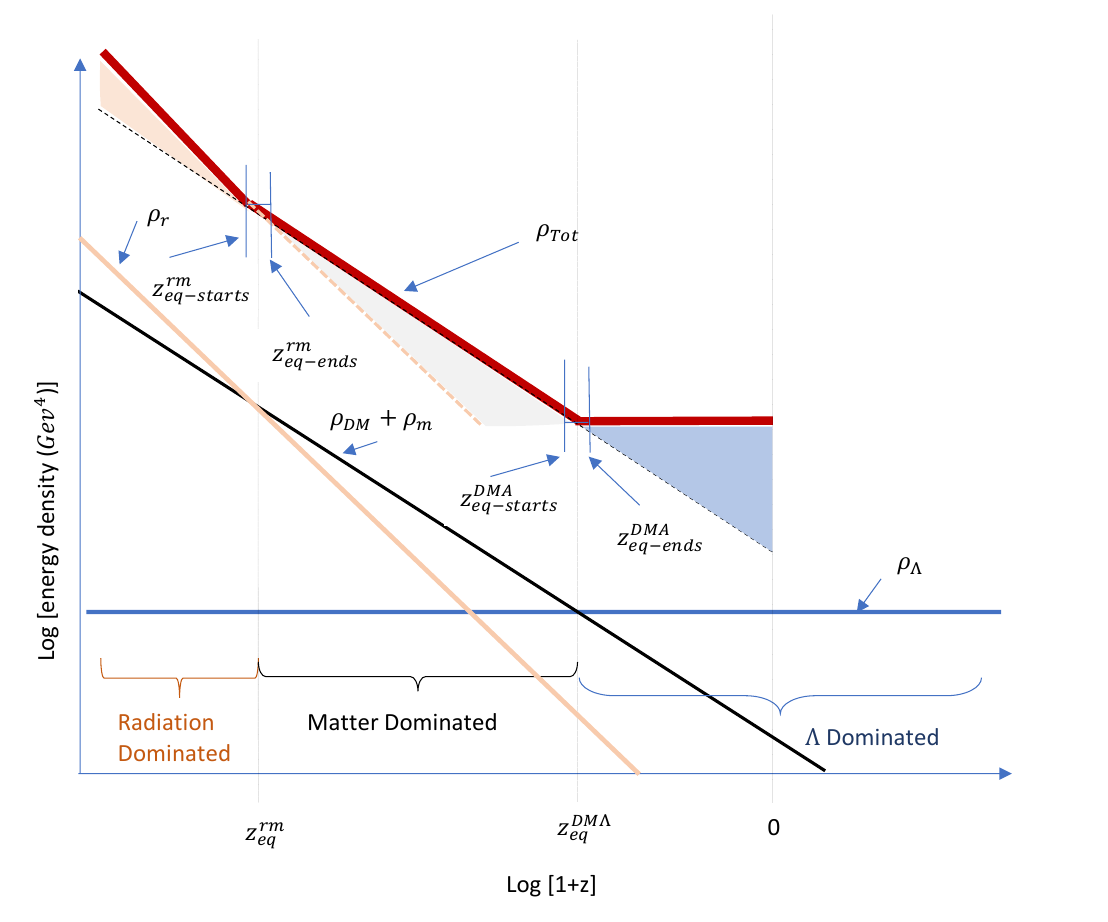}
\caption{\it The evolution Radiation ($\rho_{r}$), Matter ($\rho_{M}$) and Dark energy density. Insert is the evolution of total energy density($\rho_{Tot}$) with redshift. $rm$ refers to radiation-matter equality phase and $M\lambda$ refers to matter-$\lambda$ equality phase. Although $\vert z_{eq-end }-z_{eq-start}\vert$ for each transition period is short compared to the various epochs, the contribution of the model dynamics may not be. }
\label{fig:fig2}
\end{figure}

In this section, we examine the modifications of equation (\ref{rhoDotTot}) as a result of restrictions that characterise the different epochs. As already stated, the dynamics of the transitions between these epochs is the goal of our analysis. Physical interpretations of the results will be of interest, given the role played by energy densities in the definition of the deceleration parameter, $q$, and the evolution of the Hubble parameter, $H$, traditionally given by 
\begin{eqnarray}
q&=&\frac{1}{2}\sum_{i}\Omega_{i}(1+3\omega_{i}),\\
\frac{\dot{H}}{H^2}&=&-(1+q),
\end{eqnarray} 
respectively and where $i=r, m, DM, DDE, NDE$. 
We approach this by looking at the fractional composition of the total energy density. Until now the deceleration parameter is treated as a constant that can be evaluated by specifying the the values of the parameters in given epochs. As we will see below, our interest in the transition period will allow the deceleration parameter to be treated as a function of cosmic time i.e. $q=q(t)$. This follows from equation (\ref{eqratio00}). At face value, this may look counter-intuitive until one considers that the fractional constituent changes with the passage of cosmic time. In particular, the solution to the evolution equation of the Hubble parameter takes the form
\begin{eqnarray}
H=\frac{1}{C - \chi(t)},
\end{eqnarray}
where $\chi(t)=-\int_{1}^{t} (q(\xi)+1) d\xi$ and where $C$ is a constant of integration. It is important to note that the time dependency is on the fractional part i.e.  $\Omega_{i}=\Omega_{i}(t)$) and not the equation of state $\omega_{i}$.
In general, equations (\ref{eqratio00}) and (\ref{rhoDotTot}) yield; \begin{eqnarray}\label{evTot}
\dot{\rho}_{Tot}
&=&-\theta\left(1+\frac{1}{3}\alpha-\xi+\gamma\omega_{DM}+\delta\omega_{DDE}\right)\rho_{Tot},
\end{eqnarray} for which one can define an effective equation of state for the total mass-energy density. This takes the  following form,
\begin{eqnarray}\label{weff}
\omega_{eff}&=&\frac{1}{3}\alpha-\xi+\gamma\omega_{DM}+\delta\omega_{DDE}\nonumber\\
&=&1-\zeta,
\end{eqnarray} where the last line comes from equation (\ref{rho1}). 
It is clear from equation (\ref{eqratio00}) that although the equation of states for $DM$ and $DDE$ do not vary, $\dot{\omega}_{eff}\neq 0$. Equations (\ref{evTot}) and (\ref{weff}) form the basis of the investigation in the rest of the chapter. We note that replacing the scale factor in equation (\ref{rho1}) its redshift equivalent
\[a(t)=\frac{1}{1+z}, a_{0}=1,\] then finding the natural log of both sides of the resulting equation, using integration by parts to integrate the integral part and finally dividing through by $\ln(1+z)$ yields
\begin{eqnarray}\label{slope1}
\frac{\ln\rho_{Tot}}{\ln (1+z)}&=&\frac{3 }{\ln (1+z)}\int{\frac{\dot{z}}{1+z}\zeta dt}.
\end{eqnarray}
The right hand side of equation (\ref{slope1}) gives the slope of $\ln{\rho}~vs \ln(1+z)$ graph. It should be obvious that this term will be negative given that the redshift decreases as cosmic time increases. What follows is the analysis of $\omega_{eff}$ in the different epochs, and how it affects the right-hand side of the equation (\ref{slope1}). We pay attention to the transition between epochs where 
\begin{eqnarray}\label{slope3}
\frac{\ln\rho}{\ln (1+z)}&=&\frac{3}{\ln (1+z)}\int_{t_{eq-starts}}^{t_{eq-ends}}{\frac{\dot{z}}{1+z}\zeta dt},\\\label{slope4}
&=&\frac{3}{\ln (1+z)} [F(z_{eq-ends})-F(z_{eq-starts})],
\end{eqnarray}
where $F(z)$ is the anti-derivative. To be discussed below, are several epochs and the transition periods between any two will constitute the domain for the anti-derivative. To solve this equation detailed knowledge of the evolutionary behaviour of the various fractions during the transition phase is required.  In the absence of such information, the integral on the right-hand side of the equation (\ref{slope3}) can still be estimated using extrapolation schemes such as the Bulirsch-Stoer method or adaptive step-size Runge-Kutta \cite{Num} to achieve equation (\ref{slope4}). In this case, we use a variety of functions to construct the information during the transition. The result is presented in figure (\ref{fig:fig2}). 

We now consider qualitative details of the various epochs and transition periods. We do this in order, beginning with the radiation-dominated epoch. We have left the case of the inflation epoch and the transition between the inflation and radiation epochs for future study. It suffices to mention the unresolved disparity between the observed and predicted values of vacuum energy. In general relativity, quantum fluctuations in the early universe contribute vacuum energy that is thought to make up the cosmological constant. The calculated and the observed vacuum energy density are orders of magnitude different. Current research suggests that, when Lorentz invariance is considered, the calculated is 60 orders of magnitude bigger than the observed \cite{Wein89,Martin12}. In other words, there is a disagreement on the value of $\zeta$, but this does not prevent analyses that involve comparison to other energy densities as we will see next. The important information here is that the uncertainty is more pronounced regarding the transition to $\Lambda$ domination.

\subsection{Radiation Dominated Epoch}
In this epoch, radiation makes up at least half the total energy.  
$\alpha>\beta+\gamma+\delta+\xi$ and in particular, $0.5<\alpha\le1.$ It follows from equation (\ref{weff}) that
\begin{eqnarray}\frac{1}{6}\le\omega_{eff}-\delta\omega_{DDE}-\gamma\omega_{DM}+\xi\le\frac{1}{3}.\end{eqnarray} It is obvious that when considering radiation dominated era $\xi<<1$ ( the fraction of $\Lambda$ is negligible), $\delta<<1$ ( the fraction of $DDE$ is negligible) and $\omega_{DM}=0$. These considerations reduce the bounds of $\omega_{eff} $ to $$ \frac{1}{6}\le\omega_{eff}\le\frac{1}{3}.$$ The upper limit coincides with the value often quoted for this epoch. As previously stated, the dynamics of the epoch are dominated by the relativistic particles.  
 In terms of physics, the temperature falls sufficiently allowing the ionised material to begin to form neutral hydrogen. We know that observational astronomy is only possible from this point on. Before this, the ionised material prevents photon propagation via the Thomson scattering mechanism. It is known that this mechanism sets a limit on the redshift of observational interest to approximately $z=1000$ unless $\omega_{r}$ is very low or $\Lambda$ energy is important, matter-domination is, therefore, a good approximation to reality. The case where $\omega_{r}$ is not low or where $\Lambda$ is important will lead to the adjustment in the maximum redshift limit. In reality, the transition appears sudden but gradual in cosmic time. This suggests that there is a need for defining and examining a {\it transition-era} between radiation and matter domination epochs.
 
\subsection{\label{four}Matter-Radiation Equality Phase}
At the matter-radiation equality, \begin{eqnarray}\label{eqsix}\rho_{r}=\rho_{m}+\rho_{DM}.\end{eqnarray} Since the energy densities have different rates of growth/decay it is clear that the matter density equality does not last long, indeed starting with a radiation ratio higher than the matter ratio (radiation-domination), then equal and finally the matter ratio becoming greater than radiation (matter-domination).
In this brief epoch $\alpha=\beta+\gamma$ and in terms of fractional energy balance
\begin {eqnarray}\label{eqratio}
2\alpha+\delta+\xi=1.
\end{eqnarray} It follows that $0.25\le\alpha<0.5$ for $\delta\ne0$ (i.e non-negligible $DDE$, recall that it interacts with $DM$ but this is compensated for in equal measure in equation (\ref{eqratio})). This yields \begin{eqnarray}\frac{1}{12}\le\omega_{eff}+\delta\le\frac{1}{6}, \end{eqnarray} here too $\xi<<1$. In general, $\delta$ is comparatively small in this transition and in standard analysis taken to be negligible. However, this is a simplifying assumption where caution is advised.

\subsection{\label{Five}Matter- Dominated Epoch} This epoch began approximately 47000 years i.e. after the radiation-matter equality. The ratios of matter and radiation were first determined from observation in \cite{Perl}. However, the ratios at the onset of the matter-dominated epoch were markedly different. By matter, we mean baryonic (ordinary matter) and non-baryonic matter ( such as dark matter). These ratios are often compared to those of the other constituents. The amount of energy in the form of radiation in the universe today can be estimated using the Stefan- Boltzmann law.
\begin{eqnarray}
\rho_{photon}&=&\frac{4}{c}\sigma T^4,
\end{eqnarray}
where $c$ is the speed of light and $\sigma$ is Stefan's constant and $T$ the temperature. Assuming a black body-radiation-filled universe at a temperature of 2.7 K, one finds the energy-mass density of photons and neutrinos to be $0.4~MeV$. This is negligible when compared to $500~MeV$ which is the estimated amount of ordinary matter mass density today \cite{Roh}.
 
In theory, this epoch started with $\beta+\gamma\ge0.5$ and $\beta+\gamma\ge\alpha+\delta+\xi$ and from equation (\ref{weff}) implies
\begin{eqnarray}
\omega_{eff}
&=&\beta+\gamma-1,\nonumber\\
\end{eqnarray} and hence
$-\frac{1}{2}\le\omega_{eff}\le0.$ The upper limit is, again, what is often quoted in literature. This is a second equality era that has not been examined as much and is less written about. This is the transition from matter to $\Lambda$ domination epoch. We look at this next.
\subsection{\label{six}Matter-$\Lambda$ Equality} In this transition period
$\beta+\gamma = \delta$ so that the effective equation of state is
\begin{eqnarray}\omega_{eff}=\frac{1}{3}\alpha-\xi+\delta\omega_{DDE}+\gamma\omega_{DM}.\end{eqnarray} $\alpha<<1$ and $\omega_{DM}=0$. But $0.5\le\delta+\xi<1$ and if $\omega_{DDE}=-1$ then
$$-1<\omega_{eff}\le-\frac{1}{2}.$$ Observations suggest that at present $\delta+\xi\approx 0.70$ i.e. the fraction of dark energy (dynamical and non-dynamical).

\subsection{\label{seven} Cosmological Constant Dominated Epoch }
The cosmological constant is appealing in terms of the modelling of how the universe evolves. This is because it allows for a better agreement between theory and observation. Generically, the gravitational pull exerted by the matter in the universe slows the expansion imparted by the Big Bang. The expansion can be estimated by measurements involving supernovae. These observations seem to indicate that the universe is expanding at an accelerated rate raising the prospect of the existence of a strange form of energy that has the effect that is the opposite of the standard gravitational pull. The cosmological constant seems to satisfy the properties of such a form of energy. Nevertheless, it has not been conclusively established that dark energy is the non-dynamical cosmological constant \cite{Sol,Wein89,Sah,Pad,Cop,Shu,Bloch,Martin12}. We mention, without delving into a discussion, some of the issues related to the cosmological constant, dark energy and the expansion of the universe. They include the fine-tuning problem and the cosmic coincidence problem. For a dark energy ($DDE + \Lambda$) dominated universe, we have
\begin{eqnarray}
\dot{\rho}_{Tot}&=&-\left(1+\frac{1}{3}\alpha-\xi+\delta\omega_{DDE}+\beta\omega_{DM}\right)\theta\rho_{Tot}.
\end{eqnarray}
but $\alpha<<\xi$ and $\omega_{DM}=0$ yielding an effective equation of state of the form:
\begin{eqnarray}\omega_{eff}
&=&-\xi+\delta\omega_{DDE}.
\end{eqnarray}The bound $\dot{\rho}_{Tot}=0$ is analogous to an effective equation of state of the form $-1\le \omega_{eff}$. This implies \begin{eqnarray}\label{eqnXi}-\xi+\delta\omega_{DDE}=-1.\end{eqnarray} There are several ways to interpret equation (\ref{eqnXi})). We could follow the common approach and allow $\omega_{DDE}=-1,$ which leads to $\xi+\delta=1$ or simply that the sum of dark sector fractions makes the total mass-energy density of the universe a scenario that is yet to be reached. Secondly, we could express the equation in the form $\omega_{DDE}=(\xi-1)/\delta$ , $\delta\neq 0$, which is negative since $\xi $ and $\delta$ are positive fractions less than 1.\\
As pointed out in \cite{Linder}, the time-varying model with $\dot{\omega}>0$  behaviour, at high redshift, is different from the cosmological constant. Suppose we interpret cosmological constant as some form of dark energy, it follows that there may be other forms of dark energy that, at early times, are very different from the cosmological constant model but which quickly become dynamically negligible for $z >1$. This characteristic suggests that dark energy with time-varying EoS may exist and could be found in the low-multipole region of the CMB power spectrum as pointed out in \cite{Caldwell}.

\section{\label{DC}Discussion and Conclusion}
It is customary to express the evolution of energy densities against scale factor as is shown in the lower section of Fig.\ref{fig:fig2}. The different epochs are then separated by points of intersection of the curves representing the various energy densities. The key implication is that only one type of energy density dominates, while all others have a negligible or non-existent role in the evolution of the total energy density. In this brief analysis, we have considered the evolution of the total energy density about the relative importance of various energy density constituents. This allows for a range of possibilities for behaviour given the relative importance of each contributor. It is important to emphasise that the standard section of Fig.(\ref{fig:fig2}) (i.e. the lower section) emerges as the limiting case. We must keep in mind the following issues: (i) we have assumed that the mass-energy material content of the universe is of perfect fluid form, (ii) the universe is of the Friedmann type with $\kappa=0$, and (iii) the interacting dark sector has no noticeable effect on the evolution of the total mass-energy density. However, these assumptions can be relaxed, and the resulting system(s) of equations can be analyzed to obtain corrections to the standard model. The various ranges obtained here allow for a wide variety of matter forms, such as domain-wall with $\omega=-2/3$ \cite{Nem}. The effective equations of state for the total energy density discussed above may alter how fields evolve in the mixture. Neglecting how one component evolves will therefore lead to an over or underestimation of the field strength. Some of these assertions will be examined in the future.

Acknowledgement:
The author expresses gratitude to the University of Cape Town's NGP for financial support. 

Appendix
\section{\label{App1} Some mathematical relations and relevant calculations }
Basic formulae and step-by-step elementary calculations: Without loss of generality we write $\rho_{Tot}$ simply as $\rho$. Writing the expansion variable, $\theta$ in terms of the scale scalar, $a(t)$, and it time derivative yields:
\begin{eqnarray}
\frac{\dot{\rho}(t)}{\rho(t)}&=&-3\zeta(t)\frac{\dot{a(t)}}{a(t)},
\end{eqnarray} where we have inserted variable $t$ to remind us of the functions of this variable. This can be expressed as derivative notation as follows
\begin{eqnarray}
\frac{d \ln\rho(t)}{dt}&=&-3\zeta(t)\frac{d \ln a(t)}{dt}.
\end{eqnarray} Integration by parts yields,
\begin{eqnarray}\label{appeq1}
\ln\rho&=&-3\zeta \ln a+3\int{(\ln a)\dot{\zeta}dt}\\
\end{eqnarray} which can be expressed as 
\begin{eqnarray}
\rho&=&a^{-3\zeta}e^{3\int{(\ln a)\dot{\zeta}dt}}.
\end{eqnarray} Equation (\ref{appeq1}) can also be expressed in terms redshift as follows
\begin{eqnarray}
\ln\rho&=&3\zeta \ln (1+z)-3\int{(\ln (1+z))\dot{\zeta}dt},
\end{eqnarray} where $1+z=\frac{a_{0}}{a}$ with $a_{0}$ taken as having the value unit. 
This is a more useful form for our visualisation. Dividing this through by $\ln (1+z)$ yields
\begin{eqnarray}
\frac{\ln\rho}{\ln (1+z)}&=&3\zeta -\frac{3}{\ln (1+z)}\int{(\ln (1+z))\dot{\zeta}dt}.
\end{eqnarray} This is equivalent to
\begin{eqnarray}
\frac{\ln\rho}{\ln (1+z)}&=&\frac{3}{\ln (1+z)}\int{\frac{\dot{z}}{1+z}\zeta dt},
\end{eqnarray}  where $\dot{z}<0$.

\section{Fractional density}
From equation (\ref{rho1}) and noting that $\zeta=\zeta(t)$, then
\begin{eqnarray}
f(t)&=&\frac{4}{3}\alpha(t)+\beta(t)+\gamma(t)(1+\omega_{DM})+\delta(t)(1+\omega_{DDE}).\end{eqnarray}
In this formulation the the individual EOS for the constituents is not changing concerning time but the effective EOS will. In particular,
\begin{eqnarray}
\dot{f}&=&\frac{4}{3}\dot{\alpha}+\dot{\beta}+\dot{\gamma}+\dot{\delta}
=\frac{1}{3}\dot{\alpha}-\dot{\xi}=-\dot{\omega}_{eff}.\end{eqnarray}
This is the point of divergence from previous work in this area of research.

\end{document}